\documentclass{vgtc}                          % final (conference style)
\graphicspath{{figures/}{pictures/}{images/}{./}} % where to search for the images

\usepackage{times}                     % we use Times as the main font
\usepackage{tabu}                      % only used for the table example
\usepackage{booktabs}                  % only used for the table example
\usepackage{lipsum}                    % used to generate placeholder text
\usepackage{mwe}                       % used to generate placeholder figures
\usepackage{multirow} 
\usepackage{mathptmx}                  % use matching math font

\onlineid{0}

\vgtccategory{Research}

\vgtcinsertpkg

\title{Gamification of Immersive Cervical Rehabilitation Exercises in VR: An Exploratory Study on Chin Tuck and Range of Motion Exercises}

\author{Haitham Abdelsalam \\ %
        \scriptsize Department of Computer Science and Software Engineering \\ Concordia University%
\and Chanelle Montpetit\\ %
     \scriptsize Department of Health, Kinesiology \& Applied Physiology \\ Concordia University%
\and Arash Harirpoush\\ %
     \scriptsize Department of Computer Science and Software Engineering \\ Concordia University%
\and Maryse Fortin\\ %
     \scriptsize Department of Health, Kinesiology \& Applied Physiology \\ Concordia University%
\and Yiming Xiao\thanks{e-mail: yiming.xiao@concordia.ca}\\ %
     \scriptsize Department of Computer Science and Software Engineering \\ Concordia University}
\teaser{
  \centering
  \includegraphics[width=\linewidth]{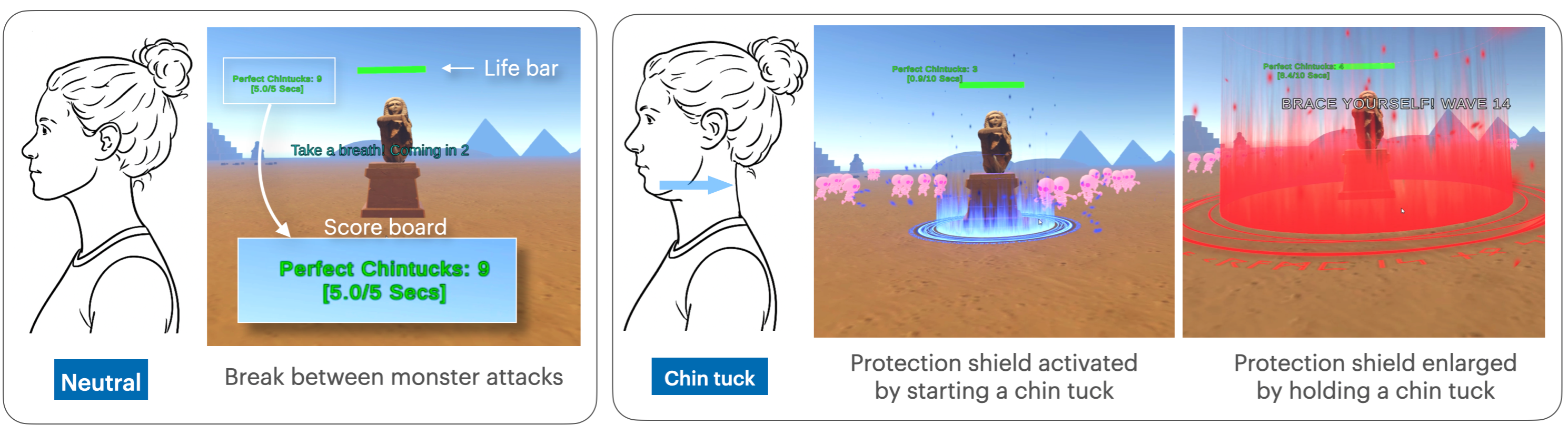}
  \caption{Overview of the Indiana Jones-themed chin tuck exercise game, with a survival and level progression strategy. The user protects the sacred artifact from monster attacks by performing chin tucks. As the user progresses to a new difficulty level, the duration of a chink tuck becomes longer with more monsters attacking. Left: When the user is in a neutral posture, the protection shield is off. Right: When the monsters attack, the user activates the protection shield by performing a chin tuck.}
  \label{chintuckdemo}
}

\abstract{
     Chronic neck pain is a prevalent condition that affects millions of individuals worldwide, causing significant individual suffering and socioeconomic burdens. Although exercise rehabilitation is a staple in relieving pain and improving muscle function for the condition, traditional one-on-one rehabilitation sessions are costly and suffer from poor adherence and accessibility for the patients. Thanks to the increasing accessibility and recent advancements in sensing and display technology, virtual reality (VR) offers the potential to tackle the challenges in traditional exercise rehabilitation, particularly through gamification. However, still in its infancy, VR-based neck exercise rehabilitation lacks exploration in effective gamification strategies and existing prototypes. To address the knowledge gap, we conduct an exploratory study on the gamification strategies for VR-based cervical rehabilitation exercises by using chin tuck and neck range of motion exercises as examples. Specifically, with different game themes, we investigate a survival and level progression strategy for muscle strengthening (chin tuck) exercise for the first time, and the suitability of ambient reward for a neck range of motion exercise. Through a preliminary user study, we assess the proposed novel VR neck rehabilitation games and they demonstrate excellent usability, engagement, and perceived health value.
} % end of abstract


\begin{document}


\firstsection{Introduction}

\maketitle

%% \section{Introduction} %for journal use above \firstsection{..} instead

Chronic neck pain is a prevalent musculoskeletal condition that affects millions of individuals globally and represent a growing socioeconomic burden, resulting in substantial costs associated with lost productivity, healthcare utilization, and long-term disability \cite{Kazeminasab2022}. Commonly associated with lack of physical activity, poor posture, and musculoskeletal imbalances, chronic neck pain often results in pain, stiffness, and limited cervical mobility \cite{Steilen2014}. Standard rehabilitation approaches, including physical and exercise therapy, have been shown to improve pain and musculoskeletal function \cite{Barreto2019}. These typically involve one-on-one sessions between the patient and the therapist with repeated sessions depending on the severity of the condition, lasting for 6$\sim$12 weeks (2$\sim$3 sessions/week). However, adherence to rehabilitation programs remains a persistent challenge due to factors, such as lack of motivation, discomfort during exercises, and insufficient engagement \cite{Himler2023,Campbell2001,Collado-Mateo2021}, leading to suboptimal outcomes. Accessibility and cost also pose as additional barriers for the patients. Thus, there is a growing need for innovative solutions that enhance engagement and adherence to improve therapeutic effectiveness in cervical spine rehabilitation.

Thanks to its increasing accessibility and recent advancements in sensing and display technology \cite{Trinidad2023}, virtual reality (VR) has demonstrated excellent promise in fitness \cite{Mouatt2020} and rehabilitation \cite{Hao2024}, by offering a stimulating environment and real-time feedback. With suitable environment design and the application of gamification strategies, VR can reduce perceived levels of exercise exertion \cite{Naugle2024} and pain \cite{Smith2020}, as well as improving enjoyment and adherence to exercise regimes \cite{Mouatt2020}, paving the way for the adoption of VR in exercise rehabilitation to tackle the retention issue in the traditional practice. Furthermore, VR technology could also potentially mitigate the cost and accessibility barriers for patients, particularly those living in remote areas. It is important to note that unlike the commonly seen exercise games (exergames), such as \textit{Supernatural} and \textit{FitXR}, which prioritize the gaming aspects that maximizes exertion, gamified physical rehabilitation must first ensure exercise protocols and goals \cite{Chiu2024}, often requiring controlled, precise, and sometimes nuanced movements. By incorporating game-like elements, such as reward mechanisms and real-time feedback, gamified exercise rehabilitation can encourage patients to actively participate in their recovery \cite{Mouatt2020,Chiu2024}. While several VR-based rehabilitation platforms exist \cite{Tosto2022,Voinescu2021}, so far, very few are tailored towards addressing the unique biomechanical and neuromuscular challenges associated with chronic neck pain \cite{Chaplin2023}. As a result, considering the growing patient population of chronic neck pain and current constraints on clinical resources \cite{Global2024}, there is an urgent need to explore effective methodologies and designs for neck exercise rehabilitation using Virtual Reality.

For chronic neck pain rehabilitation, strengthening exercises to support posture and prevent recurrence, and stretching/range of motion exercises to improve mobility are two staple types of interventions. Thus, to explore novel gamification strategies and themes in virtual reality for neck pain rehabilitation, we target these two types of exercises for our study by proposing a set of two novel VR games. These include an Indiana Jones-themed chin tuck game for muscle strengthening and a space-themed neck range of motion exercise for neck mobility. As part of the novelty for this exploratory study, we examined action game-inspired \textit{survival and level progression strategy} for the chin tuck exercise, and an \textit{ambient reward mechanism} by altering the virtual environment elements for the neck range of motion game. The goal of the study is to evaluate the usability, acceptability, and design elements of the proposed VR rehabilitation framework in healthy individuals (some with mild self-limited neck pain) before extending its application to those with chronic neck pain. \textit{We hypothesize that} the proposed game designs are user-friendly, engaging, and suitable for the intended exercises. By integrating evidence-based rehabilitation principles with VR gamification elements, we intend to cultivate an engaging, accessible, and effective solution for chronic neck pain.

\section{Related Works}
\maketitle
Virtual reality has increasingly been applied to physical rehabilitation due to its ability to provide engaging, entertaining, and controlled exercise environments. So far, most existing VR rehabilitation applications \cite{Voinescu2021} focus on the motor function of the limbs due to conditions, such as stroke, with a few recent applications for low back pain \cite{Orr2023}. In comparison, there is still a strong lack of VR applications and related explorations of prototypes and gamification strategies for the rehabilitation of chronic neck pain in both the research and commercial domains \cite{Hao2024}. Among the limited studies, Orr et al. \cite{Orr2023} conducted a retrospective rehabilitation study for low back and/or neck pain, and for neck pain, they prescribed the patients the commercial VR rehabilitation game called \textit{Rotate} from \textit{XRHealth} (www.xr.health), which formulates neck movements as controlling a dragon to collect tokens in the sky. Their study confirmed the feasibility and safety of VR-based exercise therapy. Similarly, Guo et al. \cite{Guo2024} conducted a randomized controlled trial with chronic neck pain patients using commercial VR neck rehabilitation games, which adopt the themes of shooting and token collection to train kinematic function of the neck. Their results demonstrate the benefit of VR in improving patient satisfaction, relief of symptoms, and exercise motivation. These prior investigations and software platforms have mainly focused on stretching and range of motion exercises, while strengthening exercises for the neck (e.g., chin tuck) are often overlooked as they could be trickier to gamify due to more nuanced movements \cite{Youssef2024}. In VR-based neck rehabilitation exercise games \cite{Orr2023,Guo2024,Hao2024}, the commonly seen themes of ``shoot game'' and ``tracing-based token collection'' have been staples in the literature and commercial products. Considering different characteristics of exercise types and their associated protocols (e.g., position holding time, repetition, and intensity), further exploration of the game themes and gamification strategies would be beneficial, particularly for neck strengthening exercises. In our exploratory study, by closely collaborating with certified athletic therapists (co-authors CM and MF) for exercise protocol design and VR game testing, we explore alternative game designs for chin tuck and neck range of motion exercises.

\section{Methodology and Materials}
\subsection{System Setup}

The software application was developed in Unity 2022.3.31 and uses the OpenXR backend for cross-headset compatibility. A Meta Quest 3 and a Meta Quest Pro VR headset via Quest Link were used during software development. The VR headset’s positional and rotational tracking is used for all exercises, with controller inputs for scene navigation and calibration confirmation. No external sensors or additional hardware are required, ensuring ease of deployment and reproducibility. For the user study, we used a Razer Blade 15 laptop (Intel Core i7 CPU, NVIDIA GeForce RTX 2070 GPU, and 16.0 GB RAM) and a Meta Quest Pro headset via Quest Link due to availability. We observed no lagging/frame freezing for our system.

\subsection{Chin Tuck Game}
\subsubsection{Game Theme, Design Rationale, and Workflow}
From a rehabilitation standpoint, chin tucks are essential for strengthening the deep flexor neck muscles, which are often weak in individuals with forward head posture or neck pain. In short, a chin tuck is a small, controlled movement of gently retracting the chin towards the neck (while making a double chin) without tilting the head upward or downward. Our designed exercise protocol involves progressing through three difficulty levels of chin tucks with different durations of 5 seconds, 7 seconds, and 10 seconds. At the easier levels (5s and 7s durations), the users are encouraged to perform as many chin tucks as they feel comfortable with a minimum set of five, while at the level of 10 seconds, the user should attempt to complete 10 chin tucks. 

Embedding this clinically relevant exercise in an engaging game environment can encourage repetition and adherence while promoting correct technique through real-time feedback. Our Chin Tuck game is designed around a narrative inspired by the action-packed Indiana Jones movies, where users are tasked with defending a sacred artifact from waves of monster attacks by using correct chin tucks to activate a protection shield. When the user fails to protect the artifact after a certain number of attacks, the life bar that represents the damage will be drained, thus losing the game. When the user survives the first two levels of chin tucks to reach the 10-second-duration level, performing 10 successful chin tucks at this level will win the game. Note that each wave of attacks lasts for a preset time to require withholding the chin tuck for required durations. The game theme was chosen to match the required intensity of the exercise to support user engagement by contextualizing rehabilitation exercises as part of a mission. In addition, the setup of the life bar is intended to boost the motivation of the performance while gauging the capability of the user's neck muscle strength by starting from easier to more challenging levels. Naturally, when the user's muscle strength still requires further training, the user will only be able to survive the easier levels. An overview of the game is shown in Fig. \ref{chintuckdemo}.

\subsubsection{Sensor Input and Calibration}
The proposed chin tuck game uses the positional tracking capabilities of the VR headset to detect backward head movement along the Z-axis, which is critical for identifying correct chin tuck posture. When the exercise begins, the system automatically captures the user's initial head position and orientation after a short delay to establish a reference (neutral) position. This calibration ensures that all subsequent movements are measured relative to the user's own neutral posture. Users can manually recalibrate during gameplay by pressing the `A' button, which resets the neutral head position and helps maintain tracking precision if posture shifts. The system dynamically monitors movement in real time by comparing the current head position and rotation to the calibrated neutral point. A chin tuck is considered valid if the backward movement exceeds a configurable threshold (kept the same value advised by co-author CM in the study) while lateral (X and Y) movement and head rotation remain minimal. Once a valid chin tuck is detected, a defensive shield is activated, giving users immediate audiovisual cues.

\subsubsection{Visualization and Interaction}
As shown in Fig. \ref{chintuckdemo}, when the user starts the game, they are placed in a desert in front of a sacred artifact with the pyramids in the background. On top of the artifact, the life bar and the score board with the number of perfect chin tucks and the progress of the current chin tuck posture towards the required duration. Interactions are primarily controlled via head posture, with occasional use of a controller button to recalibrate. At the beginning of the session, the system introduces the objective and calibration mechanism via animated image pop-ups. The chin tuck game leverages a wave-based gameplay loop that combines audiovisual feedback with active neck posture. When a monster wave begins, users are prompted to maintain a chin tuck posture, which activates a glowing protective shield around the artifact with a sound effect. As the user holds the chin tuck for the required duration, they earn points for `perfect' tucks, while shorter holds are recorded in the system as well. As the chin tuck holding prolongs, the protection shield will enlarge over time, accompanied by color transition from blue to red, haptic feedback via the controller, a particle system that visually scales over time, and an ambient sound effect. At different difficulty levels, the monsters are also coded with different colors to remind the users of the level they are at. Between waves, the user is informed that they can take a brief break to reduce fatigue and allow them to recalibrate the neutral posture if needed. A countdown precedes each wave, building anticipation and preparing the user mentally and physically for engagement. Throughout gameplay, visual cues, such as color changes, score updates, and audio messages, keep the user engaged and informed about their progress.  These add to the sense of urgency and achievement. Finally, an immersive audio environment plays a looping music to match the Indiana Jones theme, and a ``sad" soundtrack to indicate the loss of the game if applicable.

\subsubsection{Reward System}
Different from most rehabilitation games, the designed reward system is based on survival and level progression. Each successful chin tuck deflects an attack, while failed tucks reduce the artifact's integrity. Each level increases the required duration of the chin tuck (e.g., 5 s, 7 s, 10 s), as defined in the game configuration. Users must complete all waves within a level to progress forward or risk failure if the artifact is destroyed.

To support user motivation and rehabilitation consistency, the game differentiates between partial and perfect chin tucks. A score tracker, text prompts that inform progress, and audio cues reinforce performance. In addition, associating prolonged chin tuck with dynamic scaling, color, and sound effect changes of the activated protection shield also acts as another reward mechanism. Successfully completing a level results in motivational messages. At the highest difficulty level (i.e., 10-second hold), a winning screen is shown upon achieving 10 perfect chin tucks. Loosing the game will result in a sad music with a damaged artifact. All of these elements aim to reward both effort and accuracy, aligning with rehabilitative goals.

\subsubsection{Customization and Play Log}
To enable personalization to the user's need, basic exercise parameters, including chin tuck duration, number of waves of monster attacks, level counts, and chin tuck detection sensitivity can be defined in a configuration JSON file for the game. Additional configuration settings include the number of perfect tucks required to win and rest durations between waves. These values can be tailored to fit users' needs before each session by modifying the configuration.

The system logs session data at the end of each playthrough, including successful chin tucks per level, as well as session timestamps and durations. This data enables later review and analysis of user performance. Finally, all game interactions, such as partial/perfect chin tucks, recalibration actions, and shield activations are timestamped to build a timeline of exercise behavior. This allows clinicians to potentially analyze user fatigue, consistency, or adherence patterns.

\subsection{Range of Motion (ROM) Game}

\begin{figure*}[t]
  \centering
  \includegraphics[width=\linewidth]{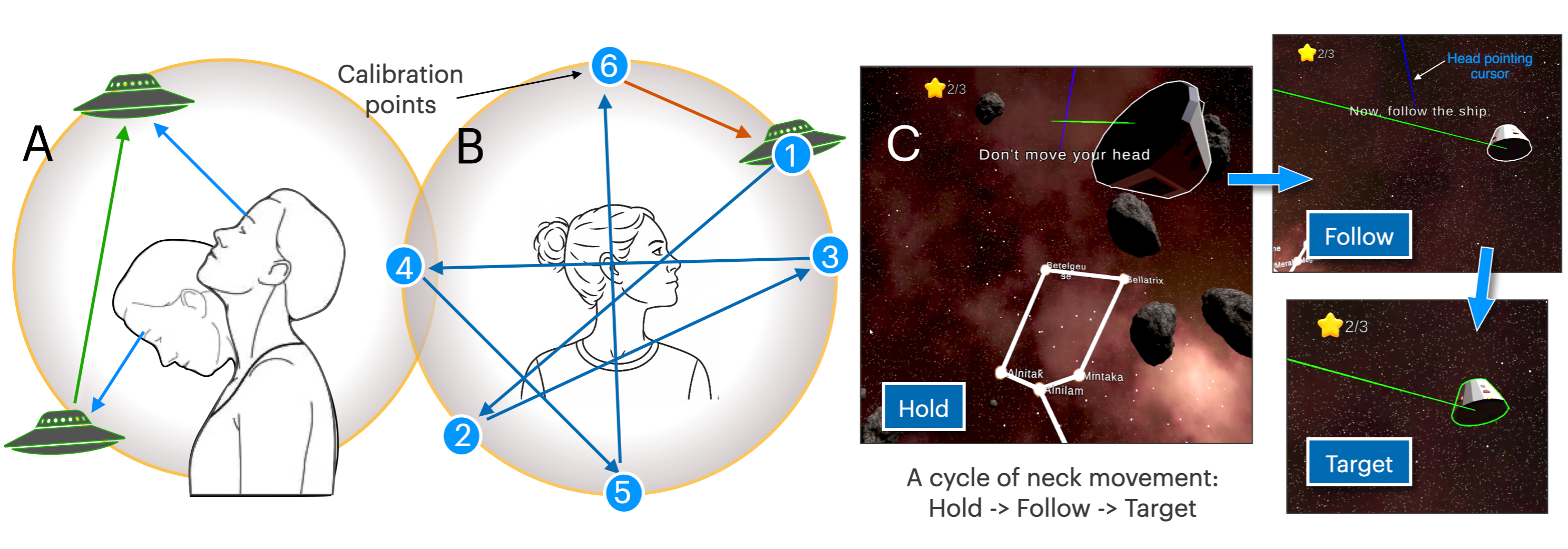} 
  \caption{Overview of the space-themed neck range of motion exercise game, where the user repeats a full set of required movement trajectories three times. A: A user performing an extension motion by following the trajectory of the spaceship with head pointing; B: Illustration of the full neck movement trajectory for one iteration, where the ordered extreme range points are obtained during the calibration stage; C: demonstration of a cycle of neck movement of holding the position, following the spaceship, and target the spaceship. Note that a blue ray emitted from the user's forehead is used to guide the motion, and once the spaceship is correctly targeted, a green halo will appear.}
  \label{rangeofmotiondemo}
\end{figure*}

\begin{figure}[t]
  \centering
  \includegraphics[width=\linewidth]{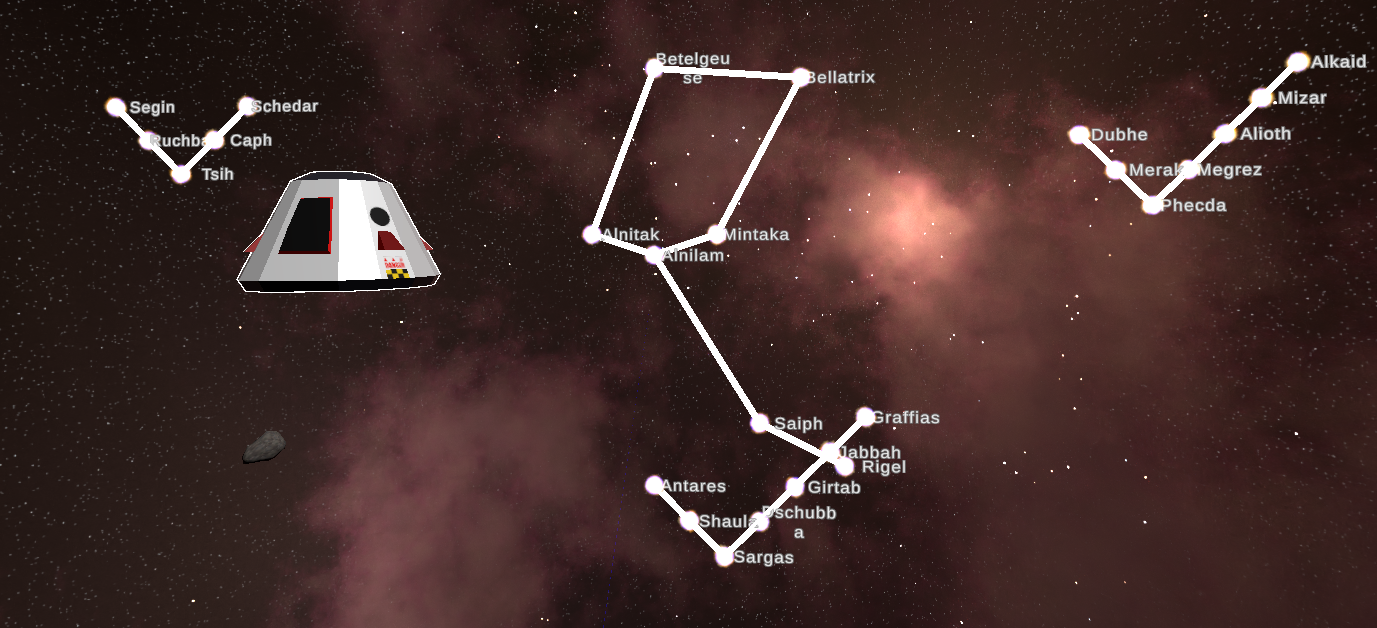} 
  \caption{Demonstration of the ambient reward in the neck range of motion game, where new constellations are drawn in the space when a set of requirement movements are completed.}
  \label{ambientreward}
\end{figure}

\subsubsection{Game Theme, Design Rationale, and Workflow}
In the context of neck rehabilitation, range of motion (ROM) exercises help restore mobility in the cervical spine. The exercise involves active moving of the neck through its natural directions of motion, including flexion (chin drop towards the chest), extension (head lift), rotation (turning the head to the side), and lateral flexion (tilt the head towards the shoulder). The ROM game features a space-themed environment, where users follow a pod-shaped spaceship by using head pointing across various individually calibrated points to enable smooth motions of flexion, extension, and rotation. As shown in Fig. \ref{rangeofmotiondemo}, we designed the exercise protocol to require the user to complete three rounds of movements (flexion, extension, and rotation), with each round following the order of the calibration points 1$\sim$6. At each point, the user will need to hold their pose first and then follow the spaceship to the next one. An additional short hold is also introduced at the mid-point between two calibration points. Upon completion, the spaceship will be removed, and the system then guides the user with an exercise illustration, prompted messages, and animation for 20 repetitions of lateral flexions alternating between the left and right.

The space theme was selected to provide a stark contrast to the chin tuck game. The sense of navigating through the stars and tracking celestial bodies provides a narrative to evoke a sense of exploration and relaxation, which aligns with the goal of increasing the neck range of motion, and gradual and steady pace of the exercise. This space exploration setting is popular and universally appealing, especially in the context of VR. Notably, unlike the more commonly seen token or coin tracing reward mechanism, we complement the space theme with a more ambient reward mechanism of  ``unlocking" hidden constellations and celestial landmarks as they perform controlled movement tasks, reinforcing both progression and a sense of discovery in a more relaxing manner. The showcase of ``unlocked" constellations is shown in Fig. \ref{ambientreward}.

\subsubsection{Sensor Input and Calibration}
For the ROM game, spatial calibration for the maximum reach of movements is conducted at the beginning of the session. The user is prompted to look in six target directions as far as possible — up, down, left, right, top-left, and bottom-right. When the user fixates on each point and confirms with a press on the `A' button, the system logs the headset’s world-space position for that orientation. These calibration points are saved in a JSON file, representing each direction's maximum reach and used as reference anchors to define the user's unique movement envelope. 

During gameplay, the spaceship moves to each of these target positions in sequence. A visual target marker (i.e., the spaceship) and a green ray trace following the target help orient the user, who must fixate on the spaceship to complete each tracing task. The system uses head orientation rather than position information from the VR headset to interact with the virtual environment, by interpreting head pointing alignment through raycasting and distance checks. Specifically, when the user maintains focus on the spaceship for a minimum configurable duration, it is counted as a successful fixation. Rather than comparing angular deviation directly, the system uses a combination of hit detection, fixation timing, and visual guiding feedback to ensure alignment.

After the flexion, extension, and rotation exercises, lateral flexion is activated. Although the physical movement involves tilting the head side to side, this is detected by measuring rotation around the headset’s Z-axis. A tilt is recorded when this rotational value exceeds a configurable threshold (kept the same value advised by co-author CM in the study). The system also monitors for a return to neutral alignment before recognizing the next tilt, ensuring clean repetitions and minimizing false positives. The system tracks successful left/right alternations based on this logic.

 The maximum angles for flexion, extension, lateral flexion (left/right), and rotation (left/right) are computed from the calibrated head positions by calculating the angle between the neutral forward direction and each target direction. These vectors are defined during calibration when the user looks in each of the six directions. These angular values represent the user's maximum achievable motion in each direction. This process ensures individualized range assessments.

\subsubsection{Visualization and Interaction}
In the ROM game, users are suspended in an immersive rendering of the space with some celestial rocks floating around to enhance the sense of depth for the environment. At the system calibration stage, text prompts are used to guide the confirmation of the maximum reach points for each direction. Upon confirmation, the point will be revealed as a sparkling star and then disappear. After completing the calibration, a spaceship will appear at Calibration Point 1 (see Fig. \ref{rangeofmotiondemo}B), and sequentially moves to various calibration-based target positions, leaving behind a green line between two position points. On top of the display, beside a yellow star, the current completed set of movements against the full three sets is displayed.

The head pointing target is shown with a blue ray emitting from the forehead of the user. When the user fixates the head pointing ray on the spaceship, a green halo will appear around it. When correctly targeting the spaceship for a required duration, it will glide toward the next position. The fixation time is configurable and differs depending on whether the target is central or peripheral. For example, fixation on the edge of the spaceship may require a longer duration (e.g., 7.5s) compared to central targets (e.g., 2s) to ensure deliberate control. For each neck movement iteration, text prompts are given to the user for the required action. Sound cues are also added to indicate successful target fixation, location transition readiness, or level completion. After completing one set of movements, the score will increment, and a new constellation will be drawn with gradual animation effects in the sky as a reward.

Upon completing three full sets of movements, users transition to the lateral flexion phase. A message prompt introduces the new task, and users are instructed to tilt their heads left and right. Each successful tilt triggers a pulse animation of the score sign and message (e.g., ``Tilted Left") along with score progression, all tracked in real time. Transitions between tilting sides are smooth and punctuated with clear message guidance.

\subsubsection{Reward System}
Progression is structured by movement set completion, with each completion of a full set contributing to the overall goal. The audiovisual feedback to confirm a successful targeting also adds to the motivation of the user. In the lateral flexion phase, users are prompted to tilt their head left and right for a predefined number of times (e.g., 10 repetitions per side). Visual and audio feedback is used to confirm successful repetitions.

Particularly, we explore a form of ambient reward mechanism by altering the virtual environment through revealing a new constellation upon completing each milestone. Unlike the commonly seen token/coin tracing reward in other rehabilitation applications \cite{Orr2023,Guo2024}, our approach intends to visually reinforce user achievement while maintaining the thematic immersion to mitigate the potential stress of missing any tokens/coins. In addition, pulse animations, level messages, and sound effects provide additional gratification. Score indicators dynamically update based on progress, and progression toward unlocking the lateral flexion exercises is clearly communicated via text prompts. These layered rewards are designed to sustain motivation while ensuring that users perform the necessary movement patterns in a clinically beneficial way. At the completion of all required movements, a celebratory animation and closing message are shown. 

\subsubsection{Customization and Play Log}
Parameters, including spaceship movement speed, fixation durations, head tilt thresholds, lateral flexion counts, and ROM movement sets are all customizable through an external configuration JSON file.  Session logs capture calibration data and the user’s maximum recorded angles in each direction, stored in JSON format for later analysis. During gameplay, key events, such as head pointing fixation success, exercise set transitions, head tilt success, and session completion are recorded. These logs support detailed performance reviews and future integration with adaptive difficulty mechanisms for personalized rehabilitation pathways.

\section{User Study and Evaluation Metrics}
Upon informed consent, we recruited 19 participants (age = 26.89 ± 3.30 years, 11 female, 8 male), who had not been clinically diagnosed with any cervical conditions or chronic neck pain (pain lasting for over 3 months), for our user study. To better understand the participant cohort, we surveyed them with a series of questions, including 1) whether they are experiencing neck issues (e.g., muscle stiffness, soreness, or pain) in the past one month, 2) the Numerical Pain Rating Scale (NPRS) in the scale of 0$\sim$10 regarding neck pain in the past 24 hours, 3) level of familiarity with VR technology, 4) past experience with exercise rehabilitation therapy, and 5) frequency of physical exercises. Among the participants, 8/19 experienced neck issues in the past one month, and in the past 24 hours prior to the study, 8/19 had a mild level of neck pain (NPRS = 1$\sim$3, mean=1.88) while one participant reported a NPRS of 4. In terms of familiarity with VR, 17/19 indicated ``Familiar” or ``Somewhat Familiar”. Out of all participants, four had received exercise rehabilitation in the past, and most of them (16/19) maintain a habit of performing physical exercises at least 1-2 times per week. No participants experienced VR sickness.

For the study, all participants were first given a brief Powerpoint presentation introducing the general knowledge of chronic neck pain and neck exercise rehabilitation, basic instructions for the developed games, and goals of the study. Following this, a tutorial was conducted to teach participants the correct movement for chin tuck exercises, and finally, a short play time for both games to familiarize them with the VR environment. All participants were asked to complete the exercises (playing the chin tuck game first) with a sitting position in a chair with a sturdy back support.

Each of the two designed games was assessed with semi-quantitative measures in the forms of user questionnaires. Specifically, the system's usability was evaluated using the System Usability Scale (SUS) \cite{brooke1996sus}, which is a Likert-scale questionnaire consisting of ten items, each with a range of 1 (strongly disagree) to 5 (strongly agree), alternating between positively and negatively worded statements. The scores for each participant are then summed, and multiplied by 2.5, resulting in a maximum SUS score of 100. A software system that receives a SUS score above 68 indicates good usability. To gain better insights into the participants' experience and perceived values of each game, we have further included an additional 1$\sim$5 Likert-scale (from strongly disagree to strongly agree) questionnaire with eleven items to evaluate perspectives, such as immersiveness, engagement, joy, difficulty of the exercises, intention of future use, and altitude towards the design elements of the games (e.g., suitability of game theme and the choice of reward/scoring mechanisms). The full list of questions are detailed in Fig. \ref{UX-chintuck}. \textit{For these questions, a higher score is more desirable.} Lastly, the participants were invited to provide free-form feedback on the positive and negative aspects of the games, along with recommendations for system improvement. 

For the total SUS score, a one-sample t-test was used to assess whether the results were significantly different from 68. For each SUS sub-score and the customized user experience (UX) questions, we compared the results to a neutral response (score = 3) to confirm the general altitudes of the participants, with the Wilcoxon signed rank test. When comparing the scores from the SUS and UX-specific questions between groups, we used the Mann-Whitney U test. Here, A p-value $<$ 0.05 was used to indicate a statistically significant difference. For the freeform comments, we summarize the insights based on the frequencies of key words (e.g., fun, engaging, easy to use, etc) used by the participants.

Lastly, besides the semi-quantitative measures, we also collected quantitative metrics from the game plays for all participants. Specifically, for the chin tuck game, we recorded the number of perfect chin tucks at different difficulty levels (5-, 7-, and 10-second hold) and the session duration. For the neck range of motion game, we recorded the maximum angles for flexion, extension, lateral flexion, and lateral rotation.

\section{Results}
%DATA and RESULTS TO BE PROCESSED

\subsection{Semi-quantitative Evaluations}
\begin{figure*}[h!]
  \centering
  \includegraphics[width=\linewidth]{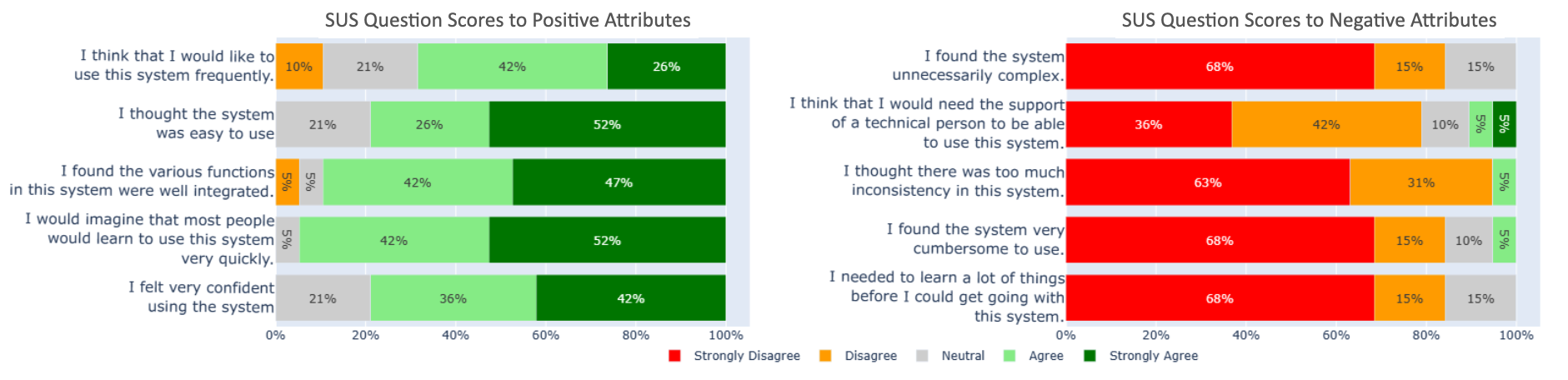} 
  \caption{Distribution of SUS scores across participants for the chin tuck game.}
  \label{SUS-chintuck}
\end{figure*}

\begin{figure*}[h!]
  \centering
  \includegraphics[width=\linewidth]{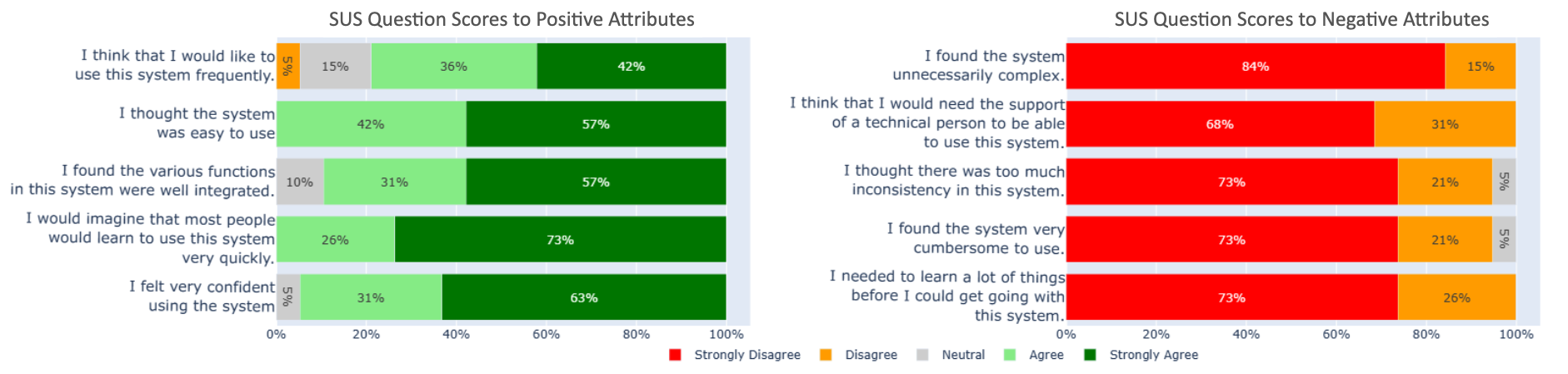} 
  \caption{Distribution of SUS scores across participants for the neck range of motion game.}
  \label{SUS-rom}
\end{figure*}

In terms of software system usability, the chin tuck game obtained an average SUS score of 83.0±16.1 while the neck ROM game achieved 90.4±9.5. Both games significantly surpassed the usability threshold of 68 (p=$7.24\times10^{-4}$ for chin tuck game and p=$5.98\times10^{-9}$ for neck range of motion game), suggesting the ease of use for the designed system, which is crucial for potential clinical adoption. When comparing the sub-cohorts with and without neck pain for the chin tuck game, although participants with pain gave a higher average SUS (86.7±16.4) than those without pain (79.8±15.9), the difference is not statistically significant (p = 0.36). For the ROM game, the SUS was also rated higher by the sub-cohort with pain (93.1±6.2 vs. 88.0±11.5) while the difference is not significant (p=0.26). Furthermore, for each game, all SUS sub-scores are significantly better than the neutral score of 3 (p $<$0.001). The detailed distributions of individual SUS question scores for each game are illustrated in Fig. \ref{SUS-chintuck} and \ref{SUS-rom}. When comparing the total SUS score between the two games, although the participants rated the ROM game more favorably on average, the difference was not statistically significant (p=0.17). When looking into the individual SUS sub-scores, the only item that differ significantly (p=0.026) between the two games is the need of technical support to use the system, with the chin tuck game receiving a score of 2.0±1.1 and neck range of motion being rated at 1.3±0.5 (a lower score is more desirable).

\begin{figure*}[h!]
  \centering
  \includegraphics[width=\linewidth]{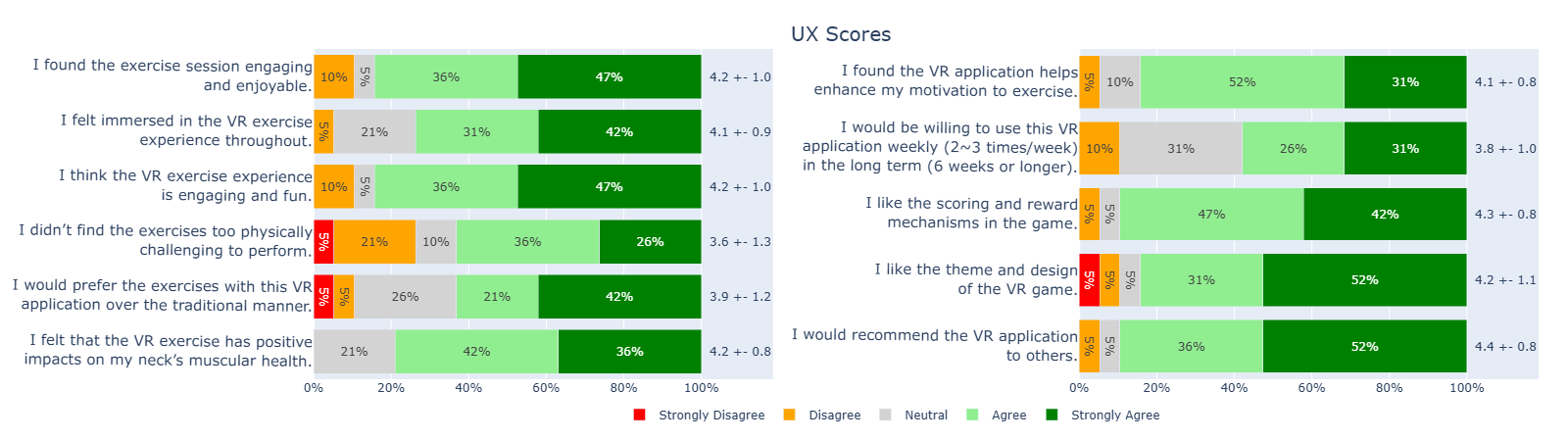} 
  \caption{Distribution of UX question scores across participants for the chin tuck game, with mean ± standard deviation displayed beside the respective bar plot.}
  \label{UX-chintuck}
\end{figure*}

\begin{figure*}[h!]
  \centering
  \includegraphics[width=\linewidth]{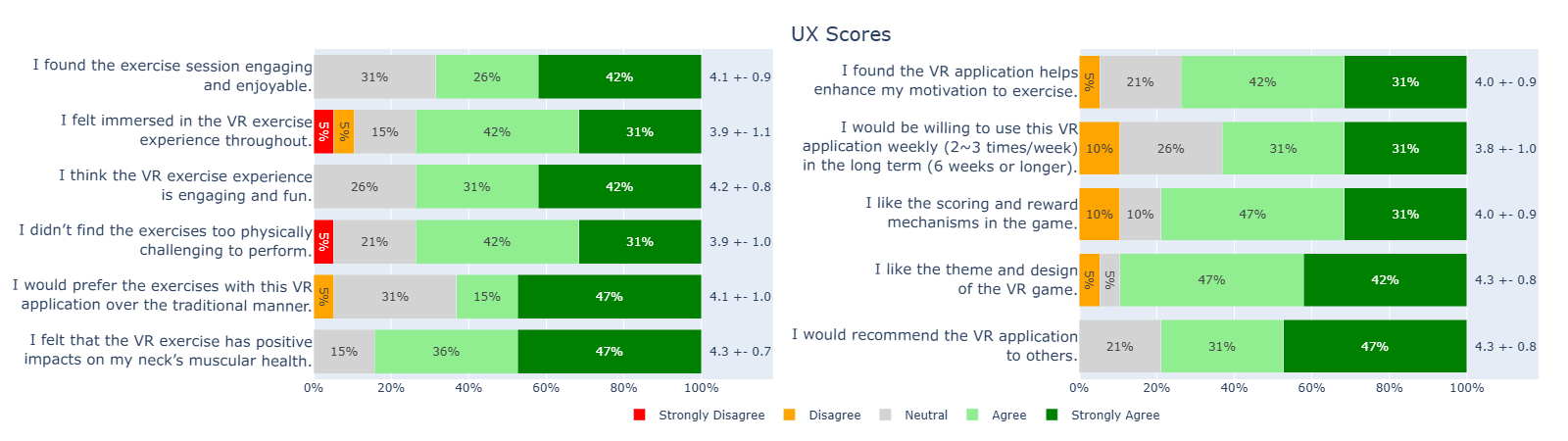} 
  \caption{Distribution of UX question scores across participants for the neck range of motion game, with mean ± standard deviation displayed beside the respective bar plot.}
  \label{UX-rom}
\end{figure*}

To complement the SUS evaluation, we surveyed the participants with additional user experience questionnaires for both games. For all the UX questions for both games, their scores are significantly better than the neutral score of 3 (p$<$0.02) except for the chin tuck game, the question related to the challenging level of the physical exercise received a score of 3.6±1.3 (p=0.062). Overall, participants found their experience with the proposed VR exercise games to be immersive, enjoyable, and engaging. In terms of exercise-related aspects, $\sim$90\% of participants gave a score of 3 or above for preferring VR exercise over the traditional manner (3.9±1.2 for chin tuck and 4.1±1.0 for range of motion) while they perceived the games to have positive impacts for the muscular health of their necks (4.2±0.8 for chin tuck and 4.3±0.7 for range of motion). Furthermore, the participants also expressed positive attitudes for using the VR games weekly (2$\sim$3 times/week) for 6 weeks or longer (3.8±1.0 for chin tuck and 3.8±1.0 for range of motion), which is common for systematic physical therapies. Finally, the design features of the games were appreciated by the participants in terms of the suitability of the game themes and scoring/reward mechanisms for the corresponding exercise types. The detailed distributions of the scores for each questionnaire item and their associated questions are presented in Fig. \ref{UX-chintuck} and \ref{UX-rom} for the chin tuck and neck range of motion games, respectively. When comparing each of the UX questions between the two games, no significant differences were observed (p$>$0.05).

\subsection{Quantitative Metrics During Gameplay}
The quantitative metrics collected during the gameplay are presented in Table 1 for both games. For the chin tuck game, 16 out of 19 participants were able to win the game by performing 10 sets of 10-second chin tucks before the life bar was drained. All recorded angles from the ROM exercises are within the expected range as verified by the athletic therapist who collaborates on the research project, and no statistically significant differences were found between the participants with and without neck pain (p$>$0.05). Additionally, the recorded angles are not correlated with the factor of sex or age (p$>$0.05), after controlling for each other.

% table for recorded quantitative results
\begin{table}[htbp]
% \centering
\renewcommand{\arraystretch}{1.5} % Increase row height for readability
\caption{Quantitative metrics recorded during game plays.} 
\label{tab:QuantitativeResults}
\resizebox{0.5\textwidth}{!}
{ % Resize table to fit text width
\begin{tabular}{|l|l|l|} 
\hline
%\multirow{\textbf{Game}} & \centering \multirow{\textbf{Metric}} & {\textbf{Result}} \\
\textbf{Game} & \textbf{Metric} & \textbf{Result} \\
%& &  \\ % Empty row for better spacing within category
\hline
%\multirow{Chin tuck} 
Chin tuck

     & \# Perfect chin tucks (5 sec) & 9.8 $\pm$ 3.8 \\
    
     & \# Perfect chin tucks (7 sec)  & 7.4 $\pm$ 3.1 \\
    
     & \# Perfect chin tucks (10 sec)  & 9.4 $\pm$ 1.5 \\
    
     & Game completion time (min) & 6.46 $\pm$ 0.73 \\
    \hline
%    \multirow{Range of Motion} 
Range of Motion
     & Max Flexion (degree) & 62.47 $\pm$ 17.58 \\
    
     & Max Extension (degree) & 49.80 $\pm$ 13.57 \\
     
     & Max Left Rotation (degree) & 45.18 $\pm$ 20.75 \\
    
     & Max Right Rotation (degree) & 44.95 $\pm$ 16.06 \\
        
     & Max Left Lateral Flexion (degree) & 43.26 $\pm$ 6.90 \\

     & Max Right Lateral Flexion (degree) & 44.36 $\pm$ 9.01 \\ 
    
     & Game completion time (min) & 8.45 $\pm$ 1.26 \\
\hline
\end{tabular}
}
\end{table}

\subsection{Freeform Questionnaire Results}
\textbf{For the chin tuck game}, 18 out of 19 participants provided comments regarding the positive and negative aspects of the system. Most of them (13/19) found their experience fun and/or engaging, and many (9/19) praised the user-friendliness and implementation. However, a few participants (2/19) also mentioned that the in-game re-calibration procedure can be difficult to use, and 3 participants found that there is a discrepancy between the duration of the required perfect chin tuck and the period of attacks from the monsters if the starting point of the protection shield is delayed. Finally, one participant suggested the implementation of a more elaborate scoreboard. \textbf{For the neck range of motion game}, all participants provided their comments for the system. Among them, 9/19 participants mentioned that the game was fun and/or engaging, and 7/19 found it easy to use. In addition, some (3/19) explicitly praised the space theme of the game. On the negative aspect, four participants found the pace of the game a bit slow, and two participants suggested additional mechanisms, such as a wait time countdown during the hold position. Lastly, a couple of users suggested additional reward mechanisms for the game. \textbf{Common to both games}, some participants suggest additional game levels to match the progress of the exercise achievements.

\section{Discussion and Future Work}

To date, there is still a great lack of systematic guidelines and software prototypes for VR/AR-based exercise rehabilitation for chronic neck pain. The proposed new VR rehabilitation system and gamification strategies demonstrate a promising framework for delivering immersive, gamified neck therapy experiences. Furthermore, we explored two different gaming strategies to accommodate the natures of the two selected exercises. For chin tucks that train deep neck muscle strength, we implemented an action-based survival and level progression strategy to encourages more repetitions of required movements \textbf{ for the first time}. For range-of-motion exercises that intend to enhance smooth, controlled joint movement, we designed a more relaxing space-themed game, with an ambient reward mechanism by altering the virtual environment upon milestones. These represent novel approaches in exercise rehabilitation VR games, and have received positive responses from the user study, where participants indicated a strong preference to use the VR games over traditional methods for exercise rehabilitation and willingness to adopt the games for long-term use.    
\\
\\
Overall, both our proposed VR games have received favorable responses from the participants in terms of ease-of-use, enjoyment, engagement, perceived benefits, and design of gamification strategies. Specifically, for the game theme and designs, the participants gave 4.2$\pm$1.1 and 4.3$\pm$0.8 out of five for the chin tuck and ROM games, respectively, and for the scoring/reward mechanisms, a score of 4.3$\pm$0.8 for the chin tuck game and 4.0$\pm$0.9 for the ROM game were given. These suggest the success of the two proposed gamification strategies. The theme and setup of the chink tuck game were especially appreciated by the participants in the freeform questionnaire. Compared with the range of motion game, the chin tuck game received a slightly lower SUS score on average (p$>$0.05), with a significant difference for the item regarding the need of professional support for the system (p$<$0.05). This is likely due to the slightly steeper learning curve and the engagement of deep neck muscles for the chin tuck. As a result, during the pre-study tutorial, we provided additional guidance to ensure that the participants learnt chin tucks correctly. In the clinical setting, we expect the patient users of the VR games to complete the first couple of sessions under the supervision of their therapists before carrying out the interventions on their own. While the ambient reward was well appreciated by the participants, one major critique about the range of motion game is regarding the idleness during posture holding at each extreme range points, which is an inherent part of the exercise protocol. To improve the game experience, we will implement additional dynamic visualization, small plays, and reward mechanisms (e.g., countdown or additional points) for these position holds in the next version of the games. 
\\
\\
As a preliminary investigation, there are still a few limitations in our study, which we intend to address in future studies. First, at the current stage, we only recruited young adults with healthy necks and self-limited mild neck pain to assess the usability and design aspects of the games while the condition is more prevalent in the mid-age group \cite{Kazeminasab2022}. After improving the games based on the received feedback, we will conduct a longitudinal study with chronic neck pain patients from a wider age range to fully assess the health benefits of the proposed system. Second, although the application is developed using OpenXR, which allows for cross-platform compatibility, we only tested the designed VR games with the Meta Quest Pro headset in the user study. We will further test the performance of the games on other VR/XR headsets in the future. Third, since previous studies \cite{Guo2024,Trinidad2023} have reported good reliability of movement tracking in existing VR headsets, we didn't perform additional accuracy validation for our application, which is out of the scope of this study. However, to ensure the clinical benefits, we will conduct related experiments to gauge the accuracy of movement tracking in different VR headsets and examine their impacts on VR-guided exercises. Finally, as mentioned by some of the participants, we didn't include multiple game levels to allow progression of exercise with theme updates in this preliminary study. During the improvement of the current system, we will investigate new paradigms, including more elaborate reward/scoring methods, and data-driven or story-telling-based exercise progression to enable a richer gaming experience and more nuanced personalization to better adapt to  various needs of the users. On this end, data collection from a diverse population with different demographics, pain levels, and underlying conditions during their game plays would allow better insights into the variations of rehabilitation progress and patient physiology. This will help automate the personalization of exercise protocols and parameters for new users of the VR rehabilitation games, potentially with machine learning algorithms.

\section{Conclusion}
To address the knowledge gap in VR-based exercise rehabilitation for chronic neck pain, we present an exploratory gamification study based on two different types of protocols, chin tuck and neck range of motion. With personalized calibration, real-time feedback, configurable parameters, and game play logs, the proposed prototypes received positive feedback from our preliminary user study with healthy participants and those with self-limited mild neck pain. Finally, the proposed gamification strategies, particularly for chin tuck exercises have demonstrated excellent feedback. As one of the few explorations in cervical rehabilitation VR, we hope the acquired insights will help guide future investigation and product building in the domain, which has a significant impact on the general population.

\acknowledgments{
This work was supported by the Fond de la Recherche en Santé du Québec (FRQS-chercheur boursier).}

\bibliographystyle{abbrv-doi}

\bibliography{template}
\end{document}